\documentclass[lettersize,journal]{IEEEtran}
\IEEEoverridecommandlockouts
\usepackage[T1]{fontenc}

\usepackage{cite}
\usepackage{amsmath,amssymb,amsfonts}
\usepackage{algorithmic}
\usepackage{graphicx}
\usepackage{textcomp}
\usepackage[table,dvipsnames]{xcolor}
\usepackage[most]{tcolorbox} %
\usepackage{url}
\usepackage{balance}
\usepackage{multirow}
\usepackage{booktabs}
\usepackage{subcaption}
\usepackage{listings}
\usepackage{array}
\usepackage{wasysym}
\newcolumntype{R}[1]{>{\raggedleft\let\newline\\\arraybackslash\hspace{0pt}}m{#1}}

\newcommand{\midsepremove}{\aboverulesep=0mm \belowrulesep=0mm}
\midsepremove
\newcommand{\midsepdefault}{\aboverulesep=0mm \belowrulesep=0mm}
\midsepdefault

\newcommand{\subsub}[1]{\noindent\textbf{#1.} }

\usepackage{tabularx}
\usepackage{tcolorbox}
\usepackage[absolute,overlay]{textpos}

\newtcolorbox{parahighlight}[1][Yellow!20]{%
  enhanced,
  breakable,
  sharp corners,
  boxrule=0pt,
  colback=#1, colframe=#1,
  left=6pt, right=6pt, top=4pt, bottom=4pt,
  before skip=6pt, after skip=6pt,
}

\newenvironment{todobox}[1][TODO]{%
  \begin{tcolorbox}[enhanced, sharp corners,
    colback=Yellow!15, colframe=Red!60!black, boxrule=0.6pt,
    title={#1}, left=6pt, right=6pt, top=4pt, bottom=4pt]}%
  {\end{tcolorbox}}
\newcommand{\todo}[2][]{%
  \begin{todobox}[{TODO\if\relax\detokenize{#1}\relax\else: #1\fi}]
  #2
  \end{todobox}
}

\def\BibTeX{{\rm B\kern-.05em{\sc i\kern-.025em b}\kern-.08em
    T\kern-.1667em\lower.7ex\hbox{E}\kern-.125emX}}

\begin{document}

\title{Ichnos+: Estimating the Carbon Footprint of Scientific Workflows Using Fitted Power Models
}

\author{
  \IEEEauthorblockN{
    Kathleen West\IEEEauthorrefmark{1},
    Youssef Moawad\IEEEauthorrefmark{1},
    Philipp Thamm\IEEEauthorrefmark{2},
    Vasilis Bountris\IEEEauthorrefmark{2},
    Giulio Attenni\IEEEauthorrefmark{3},
    Magnus Reid\IEEEauthorrefmark{1},\\
    Yehia Elkhatib\IEEEauthorrefmark{1},
    Lauritz Thamsen\IEEEauthorrefmark{1}
  }

  \IEEEauthorblockA{
    \IEEEauthorrefmark{1}\textit{University of Glasgow}, United Kingdom
  }

  \IEEEauthorblockA{
    \IEEEauthorrefmark{2} \textit{Humboldt-Universität zu Berlin}, Germany
  }

  \IEEEauthorblockA{
    \IEEEauthorrefmark{3} \textit{Sapienza University of Rome}, Italy
  }
}

\maketitle

\begin{textblock*}{\textwidth}(17.0mm,260mm)
    \begin{tcolorbox}[width=\textwidth, colback=gray!10, colframe=gray!10, sharp corners, boxrule=0.5pt, boxsep=2pt]
        \centering \small \textbf{For the purpose of open access, we have applied a Creative Commons Attribution (CC BY) license to this version of our paper.}
    \end{tcolorbox}
\end{textblock*}

\begin{abstract} 
As data-intensive scientific workflows scale to facilitate the automation of analysis of increasing amounts of data, their resource-intensive and long-running execution incurs significant energy consumption and carbon emissions. Given the already significant and rising emissions from the ICT sector, it is crucial to quantify and understand the carbon footprint of scientific workflows. However, existing tooling 
is commonly not usable in shared, virtualized environments or resorts to power models that are based on only one or two generic data points.

To address this gap, this paper presents Ichnos+, a novel system to quantify the environmental footprint of Nextflow scientific workflows. Ichnos+ enables post-hoc footprint estimation based on existing workflow traces, node-specific power models for the computational resources utilized, and carbon intensity data aligned with the execution time. We evaluate Ichnos+ against hardware-level energy measurements obtained using Intel RAPL, and the nf-core co2footprint plugin, which implements the Green Algorithms methodology. We find that Ichnos+ is capable of estimating workflow energy consumption with an estimation error of 10.8\% %
across three compute clusters, significantly outperforming the nf-core plugin. We further show that Ichnos+ extends beyond operational carbon to estimate embodied emissions as well as water and land use. Finally, we demonstrate how Ichnos+ can be extended for another workflow system, Apache Airflow, maintaining a similarly high degree of estimation accuracy.
\end{abstract}

\begin{IEEEkeywords}
scientific workflows, cluster computing, carbon footprint, energy estimation, sustainable computing
\end{IEEEkeywords}

\section{Introduction} 
Scientists in many fields, including genomics, materials science, and remote sensing, need to analyze increasing amounts of data \cite{muirRealCostSequencing2016b, fellowsyatesReproduciblePortableEfficient2021, schaarschmidtWorkflowEngineeringMaterials2021, berrimanMontageGridEnabled2004}. Scientific workflow systems facilitate the automation of such analyzes, enabling scientists to compose pipelines out of black-box tasks with data dependencies between them. 
Because these workflows are often used to process large quantities of data, they tend to be resource-intensive and long-running, leading to significant energy consumption and, therefore, carbon emissions. Indeed, the growing popularity of big data applications has been identified as a driver of the increasing emissions of the ICT sector~\cite{freitagRealClimateTransformative2021b, 11367578}.  

Scientific workflow systems such as Nextflow~\cite{ditommasoNextflowEnablesReproducible2017a} allow for the design, execution, and monitoring of workflows on heterogeneous clusters. 
While these systems usually generate detailed performance traces and logs for executed workflows, they do not produce a record of the energy consumed and carbon emitted or other connected environmental impacts. Consequently, users must manually monitor power consumption with hardware/software power meters or, otherwise, use a methodology like Cloud Carbon Footprint(CCF)\footnote{\label{ccf-footnote} \url{https://www.cloudcarbonfootprint.org/docs/methodology/}} or Green Algorithms (GA)~\cite{https://doi.org/10.1002/advs.202100707}.

In practice, monitoring power consumption requires the user to obtain physical access to attach a power meter or sufficient privileges to enable a software-based tool like Intel's Running Average Power Limit (RAPL) prior to executing a workflow. Without this step, power consumption can only be estimated based on coarse-grained utilization averages. This is possible using the CCF and GA methodologies, though at reduced accuracy.
The GA methodology relies on vendor-specified Thermal Design Power (TDP) of assigned compute resources, a proprietary metric that does not reflect key processor settings, such as processor frequency, and does not indicate idle power consumption. 
Meanwhile, the CCF methodology builds a linear power model between the power consumption measured or estimated at 0\% and 100\%; however, the approach also only uses two data points and does not reflect CPU processor settings.
In either case, to translate the energy consumed into carbon emitted, users need a measure of carbon intensity (CI), such as a yearly average or a more fine-grained metric. Generally, CI measures the amount of carbon ($CO_2e$) produced per kilowatt-hour ($kWh$) of electricity consumed, and varies across different locations, seasons, and times, depending on the sources generating electricity and the demand on the grid.

To address these limitations, we developed \emph{Ichnos+}, a trace-driven and resource-aware system for estimating the carbon footprint of Nextflow workflows. 
First, Ichnos+ takes as input the automatically generated workflow trace produced by Nextflow. The use of these traces ensures that users do not need to manually monitor power consumption, enabling the analysis of any previously executed Nextflow workflow.
Next, Ichnos+ enables users to automatically fit a device-specific, up-to-date linear power model for utilized compute resources to accurately reflect processor settings.
Finally, Ichnos+ converts the estimated energy consumption to overall carbon emissions using high-resolution time-series CI data for each workflow task and only resorts to coarse-grained yearly averages where location-specific CI time series data is not available. 
Ichnos+ also reports estimated energy consumption and carbon emissions per workflow task, providing greater granularity than existing methodologies, and allows users to identify which of their tasks have the largest footprint to address.
Given that energy consumption and compute infrastructures have an additional impact on water consumption and land usage, as well as producing embodied carbon emissions, these factors must be considered when evaluating the environmental burden of scientific workflows.
Therefore, Ichnos+ can be configured to provide estimates for workflow embodied carbon emissions, as well as water and land use. 
Building on previously presented preliminary results~\cite{west2025ichnoscarbonfootprintestimator}, this article contributes:
\begin{itemize}
    \item An extended design and implementation of Ichnos+ as open-source software\footnote{\url{https://github.com/GlasgowC3lab/ichnos}} featuring new functionality to estimate embodied carbon emissions, water and land use (\S\ref{sec:design}) from workflow traces, energy mix data, and infrastructure metrics;
    \item An expanded evaluation\footnote{\url{https://github.com/GlasgowC3lab/ichnos-results-evaluation}}, assessing the accuracy of the automated power models generated with Ichnos+ to estimate energy consumption, including a comparison of energy estimations with RAPL and the nf-co2footprint plugin as a baseline methodology using traces from six real-world Nextflow workflows (\S\ref{subsec:power-modelling} -- \S\ref{subsec:vs-nfco2footprint-plugin}), additionally showing the system's capability to quantify environmental impacts, namely operational emissions (using both average and marginal CI), embodied emissions, land use, and water use (\S\ref{sec:eval:impact});
    \item A demonstration of system extensibility beyond Nextflow to the Apache Airflow workflow system, detailing the integration process and comparing energy consumption estimates against RAPL (\S\ref{sec:eval:extend}).
\end{itemize}

\section{Background}

\subsection{Scientific Workflows}  
Scientific workflows automate data analysis processes that support a scientific objective. They are often defined in terms of their tasks and data dependencies, and are represented using directed acyclic graphs, or as pipelines.

\begin{figure}[b]
  \centering
  \includegraphics[width=0.8\linewidth]{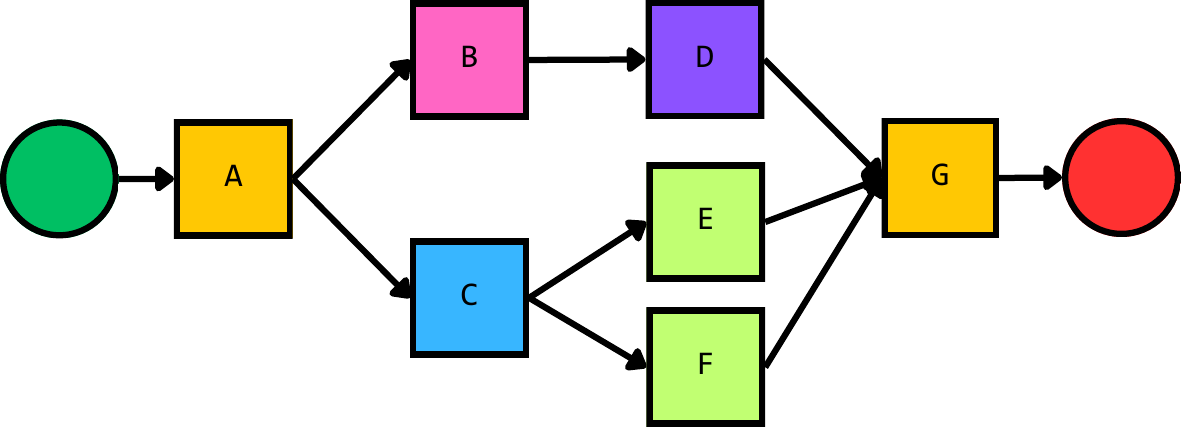}
  \caption{Simple representation of a scientific workflow, composed of seven tasks (A--G).}
  \label{fig:basic-workflow}
\end{figure} 

Figure~\ref{fig:basic-workflow} represents a scientific workflow with seven tasks (A--G) and the dependencies between them. The tasks that form a workflow are considered black-box processes, receiving input from one or more previous tasks, undergoing some processing, and producing output to send on to subsequent tasks. For example, Tasks B and C can only start after receiving input from Task A. As tasks are individual processes, both tasks B and C could run in parallel if there was sufficient capacity. 
Dedicated resource usage data could therefore be generated for each task, such as its runtime, status, and resource utilization on the compute resource it was executed on. 

Scientific Workflow Management Systems (SWMS), such as Nextflow and Pegasus, enable scientists to design, execute, and monitor workflows on heterogeneous infrastructure. A key feature of such platforms is the generation of performance traces at the individual task level. 
Similarly, general-purpose workflow systems, like Apache Airflow, operate pipelines formed of interdependent tasks, though not necessarily designed for scientific applications.

\subsection{Carbon Intensity Signals}
The carbon intensity (CI) of electricity measures carbon dioxide equivalent released per unit of energy produced. 
Generating energy from renewable energy sources such as wind or solar reduces CI. CI also typically decreases during periods of time when the demand on the grid is low. 
In this work, we use grams of carbon dioxide equivalent per kilowatt-hour ($gCO_{2}e/kWh$) to measure CI. 

CI can be quantified using two signals: average and marginal. 
Average reflects the overall grid emissions at the time when electricity is requested, factoring in each energy source's relative share and emission rate. 
Marginal measures the emissions of the specific energy sources used to meet additional load at the time when electricity is requested. 
While marginal CI is in theory preferable for measuring the impact of demand shifting, average CI is more readily available and used for emissions accounting.
Moreover, average and marginal CI are available at different levels of granularity, such as yearly averages, hourly values, and 5/15/30-minute values. 
Given that grid energy mixes are changing continuously, we recommend that users select the most granular time and location specific CI signal, whether that is average or marginal.

In addition to using CI signals to quantify the operational carbon emissions, the embodied carbon emissions are also often significant and must, consequently, be understood as well. 
Embodied emissions refer to all emissions associated with the manufacturing and life-cycle of physical compute hardware, including end-of-life activities like device disposal. These emissions are typically estimated as part of a Life Cycle Assessment (LCA)~\cite{LCACritique1995}, and measured in \(\mathrm{kgCO_2e}\).
The embodied carbon emissions for a certain compute workload can be attributed by using the ratio of a workload's runtime and the device's expected lifetime, which we take to be 4 years for CPUs, as per the CCF methodology\textsuperscript{\ref{ccf-footnote}}. 
However, LCA data on compute devices are not always available and the device's expected lifetime can also only be estimated. 

\subsection{Land Usage and Water Intensity Signals}
In order to reflect the water and land usage of executing a workflow, we must consider both the water and land required to generate the electricity consumed by the workflow execution as well as the water and land needed by the data center where the workflow was executed.

Over time, the electricity generated by a grid is produced by a changing mix of energy sources, all of which have different energy water intensity factors (EWIF) and energy land usage intensity factors (ELIF). Each source can be retrieved from data on an electricity grid's energy mix, allowing to factor in the relevant EWIF and ELIF factors for each source proportionally while computing the average water and land intensity factors over time, measured in liters of water per kilowatt-hour ($l/kWh$) and square meters per kilowatt-hour ($m^2/kWh$), respectively. 

Furthermore, the data center where workflows are executed exhibits its own water and land footprints~\cite{attenni2025}. 
The Water Usage Effectiveness (WUE) factor is defined as the ratio of the water consumed by the data center to its IT energy consumption, expressed in liters per kilowatt-hour ($l/kWh$).
Similarly, the Land Usage Effectiveness (LUE) factor represents the ratio of total land occupied by the facility to its IT energy consumption, expressed in square meters per kilowatt-hour ($m^2/kWh$). 

\section{System Design}
\label{sec:design}
We now discuss the requirements for the design of Ichnos+ and provide an overview of the estimator system's design.

\subsection{Requirements}
We identify the following requirements from which we derive the design of Ichnos.

    \paragraph{Enable post-hoc estimation} Ichnos+ enables footprint estimation after workflows have been executed. This lets users analyze the carbon footprint of previous \textbf{and} new experiments, from executions that may have occurred on local devices, clusters, and cloud infrastructure. Given this, users may no longer have access to the infrastructure used to execute workflows, whether the devices were replaced, or only temporarily booked in a cloud environment.

    \paragraph{Use resource utilization data} Resource utilization monitoring is often more readily available than power monitoring.  
    Users, for example, typically lack access to power monitoring tools in shared cloud environments. Also, users may not have configured power monitoring when workflows were executed, but will often still have access to monitoring data. Hence, Ichnos+ uses existing workflow traces that contain task-level resource usage data, without energy measurements.  

    \paragraph{Estimate CPU and Memory Energy Consumption} Ichnos+ focuses on the energy consumption of CPU and memory, as these components typically exhibit the widest dynamic power consumption ranges attributable to specific load on compute resources~\cite{barroso2025energy}. We focus on CPUs and do not consider GPUs, since workflow systems like Nextflow typically focus on executions on CPUs. In addition, workflows are often executed on shared and distributed resources, where storage access may not be limited to the nodes on which workflow tasks are executed, yet workflow traces \emph{are} commonly limited to resource usage by tasks on specific nodes.

    \paragraph{Estimate Operational and Embodied Carbon Emissions} Ichnos+ estimates both the operational and embodied carbon emissions. By default, it will make use of electricity grid emissions data like the average and marginal CI for operational emission estimations. In contrast, users \emph{can} choose to also estimate embodied emissions, with hardware LCAs detailing embodied carbon emissions less widely available than CI data.

    \paragraph{Estimate Workflow Carbon Footprint} Many workflow traces are produced for individual workflow runs and offer no insight into other processes running on the same machines. We, therefore, rely on the generated workflow traces for accounting for the correct shares of an overall system.

    \paragraph{Estimate Water and Land Use} 
    Once workflow energy consumption has been estimated, users can choose to also estimate water and land use by configuring the respective intensity factors, based on the energy mix in the utilized region.

    \paragraph{Extensibility} 
    While Ichnos+ was developed to estimate the environmental footprint of Nextflow workflows, the system is intended to be extensible, allowing for the footprint of applications implemented using other workflow systems to be estimated. Though, resource utilization monitoring is required for systems that do not log this information themselves. We provide functionality to integrate other systems and demonstrate how Ichnos+ can be extended, using the example of Apache Airflow workflows.

\subsection{System Overview}
\begin{figure}[htbp]
  \centering
  \includegraphics[width=1\linewidth,trim=0 0 0 0.7cm,clip]{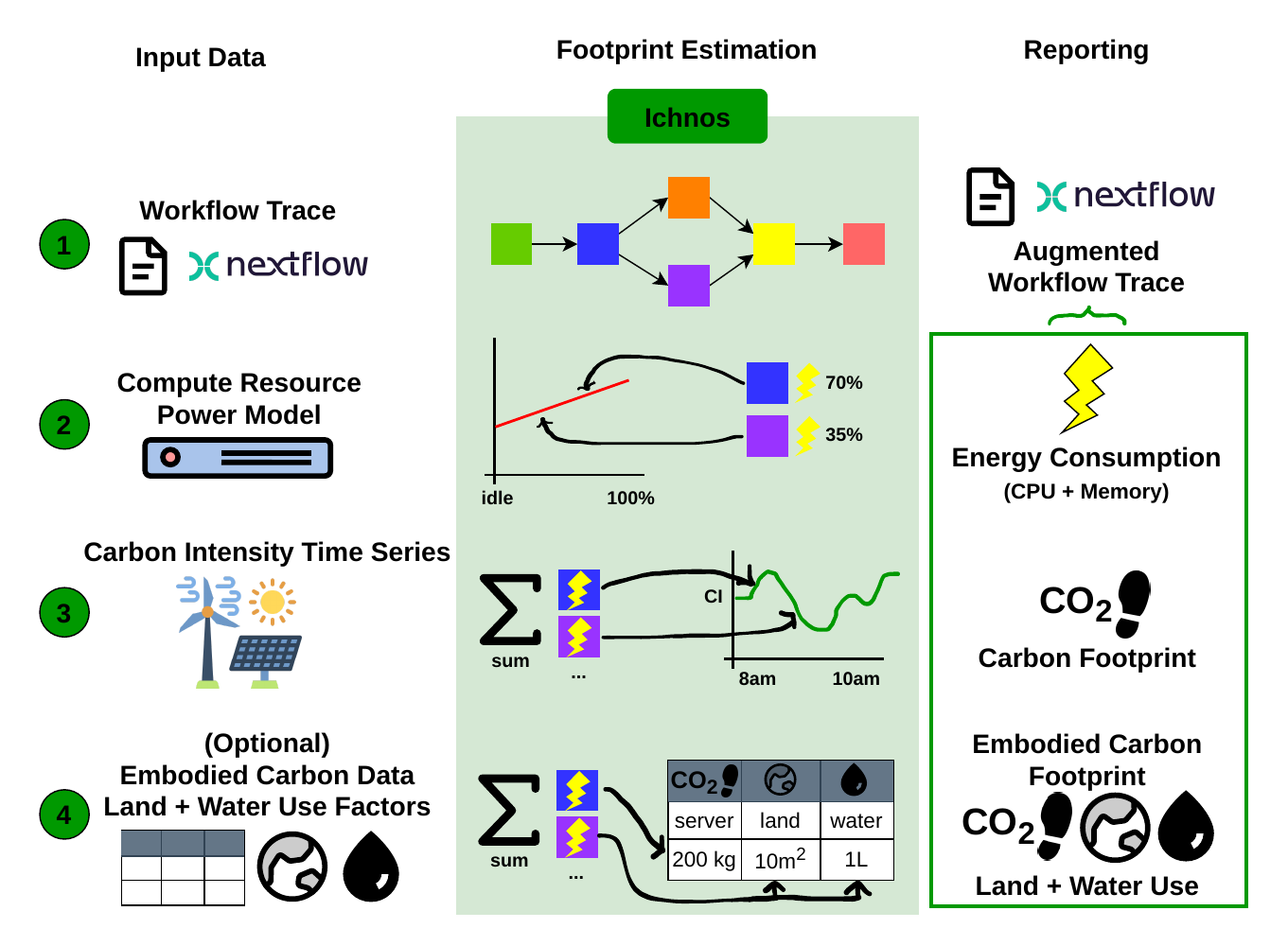}
  \caption{High-level design of the Ichnos+ Carbon Footprint estimator system with per-task power and emissions estimation, based on provided input data, and detailed impact reporting.}
  \label{fig:design}
\end{figure}

Ichnos+ (Figure~\ref{fig:design}) is a system that produces an estimate of the operational end embodied carbon emissions from the execution trace of a Nextflow scientific workflow using power and energy data aligned with the execution. Ichnos+ can also be used to calculate estimates for water and land use. 

\noindent First, the user must provide certain \textit{Input Data}:
\begin{enumerate}
    \item A workflow trace containing a task-level summary of resource usage including execution runtime, CPU utilization, and allocated memory;
    \item The power model selected to estimate the power consumption, which can be either an automatically generated power function or a regression-based model configured to reflect processor settings; and
    \item CI data supplied as fine-grained time-series data, if available, or as a coarse-grained average.
    \item (Optional) The embodied carbon emissions of utilized compute and memory resources, alongside the water and land use factors specific to the region where the workflow was executed.
\end{enumerate}

Next, during the \textit{Footprint Estimation} phase, resource usage data are extracted from the workflow trace for each task, and the energy consumption is estimated using the selected power model. Subsequently, the energy consumption per task is translated into carbon emissions using the provided CI data. This estimates operational carbon emissions by aligning the tasks of potentially long-running workflow applications with CI data matching the specific execution times. These estimations are summed to calculate the power consumption and carbon emissions for the overall workflow execution. 

Finally, during the \textit{Reporting} phase, the energy consumption and carbon emissions estimated for each task are summarized in a trace file, alongside a summary of the overall carbon footprint. 
When enabled, the embodied carbon emissions, as well as water and land use are also reported in Ichnos+ summary file, allowing users to better understand the environmental impact of their workflow executions.
We also identify the 10 most energy-intensive and the 10 longest-running workflow tasks, allowing users to review their relative emissions, and to consider the potential of aligning tasks with fluctuating CI -- applying existing carbon-aware methods to reduce the overall footprint~\cite{wiesner2021let, hanafy2023carbonscaler, west2026systematic}.

\subsection{Automated Power Modelling}
\label{subsec:power-modelling-explanation}
\begin{figure}[htbp]
  \centering
  \includegraphics[width=1\linewidth]{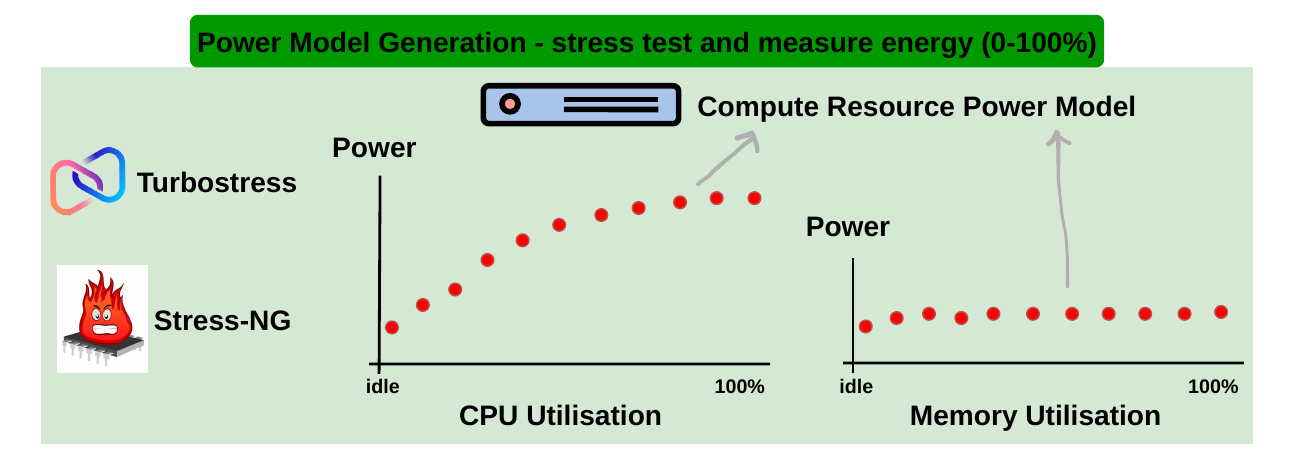}
  \caption{Overview of automated power modelling in Ichnos+.}
  \label{fig:design-power-modelling}
\end{figure}

Ichnos+ supports generating power models for utilized compute resources in an automated manner, selecting the most accurate available model based on automated readings over different levels of resource utilization as depicted in Figure~\ref{fig:design-power-modelling}.

\subsub{Approach}
The estimator system contains scripts which are used to take node-specific power measurements. First, several measurements of the energy consumed by the CPU are obtained by stress-testing the CPU using the Turbostress\footnote{\url{https://github.com/teads/turbostress}\label{turbo}} tool. Ichnos+ has a default setting of 11 measurements, adjusting CPU load from 0\% (idle) to 100\% (max) in 10\% increments. These measurements should ideally be taken at the time of execution, on the compute resources where workflows were executed, so they reflect the current processor settings such as the governor selected (which decides how the CPU frequency is adjusted based on CPU demand) on each node.
Next, similar measurements of the energy consumed by memory are taken using the stress-ng tool\footnote{\url{https://github.com/ColinIanKing/stress-ng}}, varying memory load from 0\% (idle) to 100\% in 10\% increments like CPU. Typically, memory energy consumption does not increase linearly as the load increases, and, instead, the power draw over time is relatively constant when load is applied, only differing between the readings at 0\% and with load~\cite{barroso2025energy}. 
Once the CPU and memory have both been measured, a power model is generated by fitting a linear model from the CPU energy consumption readings, while a constant is taken from the memory readings, averaging over the measurements taken with load. 
The power modeling phase should be repeated at regular intervals through compute resource lifetimes as well as when hardware is changed, to account for altered device performance. We evaluate the accuracy of Ichnos+'s power model generation in Section~\ref{subsec:power-modelling}.

In the scenario where users estimate energy consumed by a historical workflow, executed on resources that they no longer have access to, or on public cloud resources, or anywhere else where a user cannot execute the energy measurements script -- Ichnos+ has the fallback option of using a linear power model, or if only the CPU model is known, a per-core value based on vendor-specified TDP. Both these fallbacks are used in existing estimation methodologies, like GA~\cite{https://doi.org/10.1002/advs.202100707} and CCF\textsuperscript{\ref{ccf-footnote}}, and can offer ballpark estimates of the energy consumption to then translate into carbon emissions. 
Similarly, the memory energy consumption coefficient in Ichnos+ can default to a constant conversion factor, as used in existing estimation methodologies.

\subsub{Estimation of CPU Energy Consumption}
When estimating the CPU energy consumption, Ichnos+ uses a linear power model in the form:
\begin{equation}
    P = \nabla_{node} (CPU_{usage}) + idle_{node}
\end{equation}

Nextflow scientific workflows can run workflow tasks on multiple nodes, with many nodes running tasks in parallel on the same node.
When using a linear power model, each task is responsible for their own dynamic power consumption, that is, their share of the over all system CPU usage ($CPU_{usage}$), which is calculated for each task by Ichnos. 
However, when tasks run in parallel, they are all partially responsible for idle power consumption, using the $idle_{node}$ value -- this must only be added once, over the periods of time when a node is running one or more tasks. 
In our work, we assume that users have reserved the compute nodes where the workflow was executed, so they would be responsible for the idle system load. If this was not the case and users had, for instance, access to 50\% of the resources of a node, they could adjust the $idle_{node}$ value to be half of the system's idle power consumption. 

Therefore, the overall CPU energy consumed by an individual compute node is, where t represents the time:
\begin{equation}
    E = \sum_{tasks}\nabla_{node} (CPU_{usage}) * t_{task} + idle_{node} * t_{node}
\end{equation}
where we sum each task's dynamic energy consumption, with the idle energy consumed by the node when active.

This approach requires a reliable estimation of the average CPU usage for each task.
We use the CPU utilization as reported by Nextflow.
Nextflow calculates the average CPU utilization of a physical workflow task by reading the total CPU time used by all waited-for child processes from \texttt{/proc/\$pid/stat} and dividing it by the total CPU time used by all processes during task execution\footnote{https://github.com/nextflow-io/nextflow/blob/master/modules/nextflow/src/\\main/resources/nextflow/executor/command-trace.txt}.
Since the calculation is based directly on data derived from system performance counters provided by Linux, the accuracy of the resulting CPU usage estimation depends on the accuracy of these counters.
In practice, prior research shows that Linux hardware performance counters are typically very accurate, with reported observed errors of less than 0.002\% across multiple workloads and CPU models~\cite{weaver_can_2008}.

\subsub{Estimation of Memory Energy Consumption}
When estimating memory energy consumption, Ichnos+ uses a constant, measured, coefficient:
\begin{equation}
E = t * mem_{size} * mem_{coeff}
\end{equation}
Specifically, memory energy consumption can be estimated in two ways.

To provide task-level estimates of memory energy consumption, we set $mem_{size}$ as the amount of memory allocated to a task, and set $t$ as the task's runtime. This approach provides granular task-level estimates, but it can underestimate node-level memory energy for CPU-intensive workflows in which memory is not well utilized, e.g. only 30\% of available memory is allocated to the workflow. Since the power consumed by the node memory is fairly constant, regardless of the exact overall memory utilization, the sum of task memory energy consumption will be significantly less than overall node memory energy consumption in these situations. 
We also note that RAPL readings of memory energy consumption are for the entire node, regardless of whether it is shared or not, meaning that comparing node memory energy consumption would provide a better comparison with RAPL. 

For this reason, Ichnos also provides a node-level estimate of memory energy consumption, setting $mem_{size}$ to the amount of memory available on the node, and set $t$ as the time that a node is running one or more tasks. This approach captures the node memory consumption regardless of how heavily it is utilized. However, this means that individual tasks that are more memory-intensive cannot be identified. 

Ichnos+, consequently, reports both of these values, with task-level estimates provided in the output trace file, and the summed task estimates in the summary file. The summary file also contains the total node-level estimate of memory energy consumption, allowing for the user to understand their overall memory energy consumption, and to identify memory-intensive tasks.

\subsection{Environmental Impact Reporting}
Ichnos+ offers human-readable files to quickly understand the footprint of a workflow for general reporting. It also produces computer-readable files as an augmented trace file where the energy consumption and carbon emissions of all tasks are reported. This is provided to enable scientists to better understand their workflow's footprint, and identify the heavy-hitting tasks that may be disproportionately contributing to the overall footprint.

\subsection{Extensibility of Ichnos+ for Other Workflow Systems}
While the Ichnos+ estimator system was developed for the Nextflow SWMS, we believe that our approach can be applied to other workflow systems.

For this, we must have the following information available for the workflow execution to make estimations:
\begin{itemize}
    \item at task-level: the runtime, CPU utilization, allocated no. of cores, and the amount of memory assigned to each task, and the compute node utilized
    \item at node-level: the generated power model, measured memory coefficient, the no. of CPU cores and total available memory, and details of the CPU model and RAM utilized if estimating embodied carbon emissions
    \item at cluster-level: the region where cluster nodes are located
\end{itemize}
With all of the required information, an Airflow workflow trace can be created, reporting this information for every individual task. Given that this is not automatically produced by Airflow, a provenance management system, HyProv~\cite{bountris2025hyprovhybridprovenancemanagement}, was used as an additional system to create the traces needed. 

We then developed a template script to convert all Airflow task records into Ichnos+ trace records, \texttt{AirflowTraceToIchnos}\footnote{\url{https://github.com/GlasgowC3lab/ichnos/blob/main/src/scripts/AirflowTraceToIchnos.py}}. This script can be used as a basis for other systems when converting from one trace format to another.

\section{Evaluation}
\label{sec:evaluation}

\noindent In this section, we present:
\begin{itemize}
    \item The experimental setup that we used to evaluate Ichnos;
    \item An analysis of the accuracy of the system's generated power models for estimating power consumed during workflow execution;
    \item An analysis of the accuracy of workflow energy consumption estimations, compared against ground truth energy measurements and a baseline, the nf-co2footprint plugin\footnote{\label{foot:nfco2}\url{https://github.com/nextflow-io/nf-co2footprint}};
    \item An analysis of the accuracy of workflow energy consumption estimations for the Airflow workflow system, comparing against ground truth energy measurements;
    \item A demonstration of Ichnos+ being used to estimate the embodied carbon, water, and land impacts of workflow executions.
\end{itemize}

\subsection{Experimental Setup}
\label{subsec:experiment-setup}
We use ground truth data, a baseline, multiple workflows and infrastructures, as well as water and land use factors for our experiments as described here.

\subsub{Comparison with Ground Truth and Baseline Method}
To evaluate the accuracy of our approach when estimating energy consumption we compare all estimates against hardware-level measurements taken using Intel's RAPL, which we consider to be the ground truth throughout our experiments.

We additionally evaluate our approach by comparing against the nf-co2footprint plugin, which implements the GA methodology~\cite{https://doi.org/10.1002/advs.202100707}. GA uses the manufacturer-specified processor TDP to estimate per-core energy consumption. The nf-co2footprint plugin has been accepted by the nf-core community and is the official Nextflow plugin for estimating carbon emissions.

\subsub{Workflows}
To evaluate our system, we used six real-world workflows from the nf-core repository\footnote{\url{https://github.com/nf-core}}, a community-curated collection of workflows implemented using Nextflow~\cite{ewels2020nf}. We selected the Atac-Seq, Chip-Seq, Nano-Seq, RNA-Seq and Sarek bioinformatics workflows, all of which rank within the top 10 most popular workflows in nf-core, to represent typical domain usage of Nextflow. We also selected Rangeland to represent another scientific domain, namely earth observation. We manually executed the workflows to produce experimental data with RAPL energy consumption measurements.%

\subsub{Infrastructure}
Throughout our evaluation, we use the following compute nodes, which are described in Table~\ref{table:compute-nodes} and are part of three clusters.
In particular, the experiments in Sections~\ref{subsec:vs-rapl} and~\ref{subsec:vs-nfco2footprint-plugin} set up Kubernetes clusters on these compute nodes and use a Ceph PVC to run Nextflow.

\begin{table}[tb]
\caption{The compute nodes used in the study.}
\label{table:compute-nodes}
\centering
\resizebox{0.95\columnwidth}{!}{
\begin{tabular}{llccl}
\toprule
& & Cores & Memory & \\
Node & Hardware & (\#) & (GB) & Type \\
\toprule
Glasgow-1* & Intel Xeon E5-2640 v2 (x2) & 32 & 64 & Cluster \\ %
Glasgow-22 & Intel Xeon Gold 6426Y & 64 & 128 & Cluster \\ %
Berlin & Intel Xeon Silver 4314 & 32 & 256 & Cluster \\ %
Dublin & Intel Xeon Platinum 8275CL & 96 & 192 & AWS \\ %
\bottomrule
\end{tabular}
}
\end{table}

\subsub{Carbon Intensity Factors}
We retrieve average CI data from ElectricityMaps\footnote{\url{https://www.electricitymaps.com/}\label{foot:emaps}}, marginal CI data from WattTime\footnote{\url{https://watttime.org/}\label{foot:watttime}} and LCA factors for embodied carbon from the Boavizta API~\cite{Boavizta2025}.

\subsub{Water and Land Use Factors}
We retrieve historical data of the electricity grid's energy mix for the regions where the compute clusters are located from ElectricityMaps\textsuperscript{\ref{foot:emaps}}. We combined these with carbon intensity coefficients taken from the IPCC~\cite{ipcc2014_ar5_annexIII} report and~\cite{owid_safest_sources}, EWIF coefficients from NREL~\cite{nrel2011}, and ELIF coefficients from~\cite{lovering2022}.
To estimate the water and land used by the data center, we must configure the WUE and LUE. These values might be known for larger data centers (such as AWS ones\footnote{\url{https://sustainability.aboutamazon.com/products-services/aws-cloud}}), but are often unknown for smaller data centers. Therefore, we exemplify making a best guess, using a WUE of 1$l/kWh$, and a LUE of 5$m^2/kWh$.

\subsection{Power Modelling Accuracy}
\label{subsec:power-modelling}
We discuss the accuracy of Ichnos+'s generated power models, in relation to the actual energy consumption.

\subsub{Power Model Accuracy}
For each available compute device, we took power consumption readings, as detailed in Section~\ref{subsec:power-modelling-explanation} for the CPU. We used these readings to generate linear and cubic regression models of consumed energy.

\begin{figure}[t]
  \centering
  \includegraphics[width=0.8\linewidth]{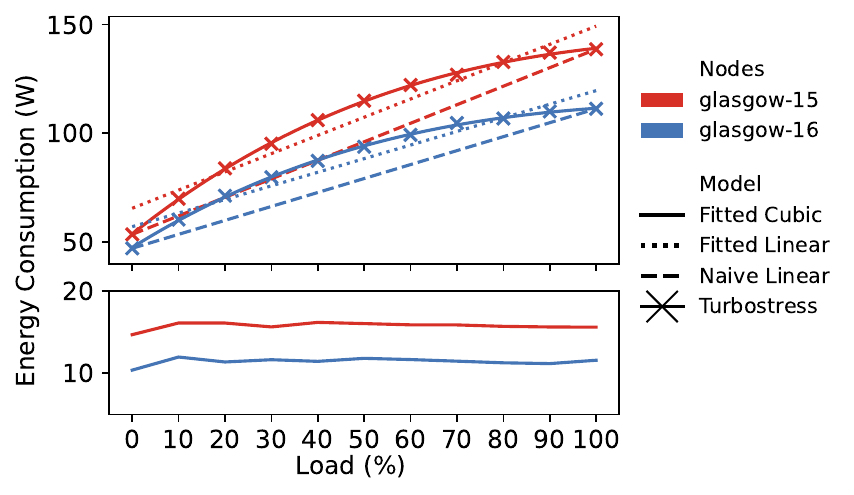}
  \caption{Power consumption versus relative load for Glasgow compute nodes utilizing the ondemand CPU governor.}
  \label{fig:compare-governor-energy-gu}
\end{figure}

\begin{figure*}[t]
  \centering
  \includegraphics[width=\linewidth]{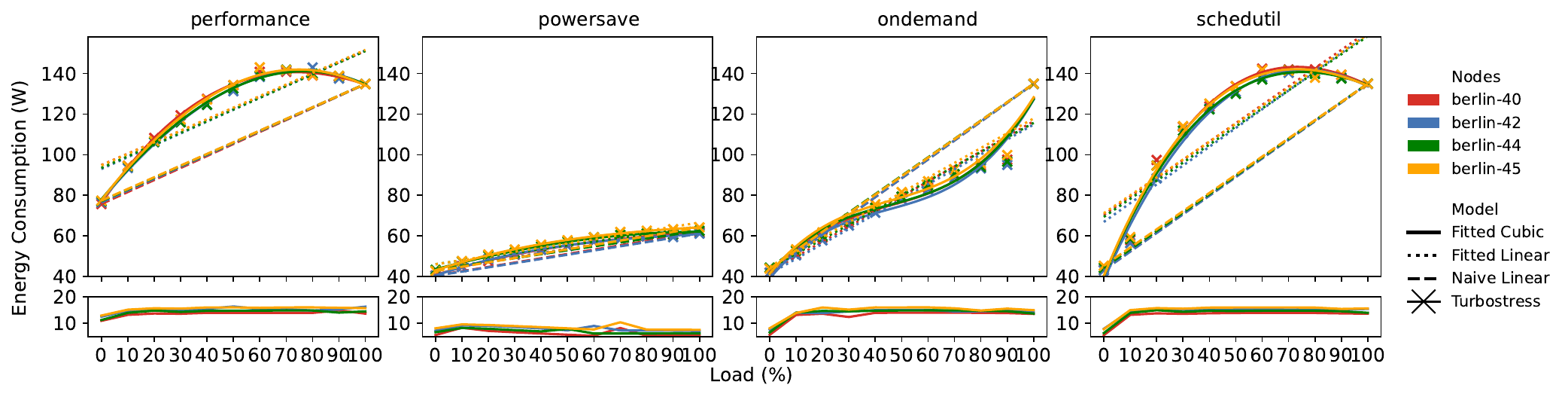}
  \caption{Power consumption versus relative load for Berlin cluster nodes utilizing different CPU governors.}
  \label{fig:compare-governor-energy-hu}
\end{figure*}

In Figures~\ref{fig:compare-governor-energy-gu} and ~\ref{fig:compare-governor-energy-hu}, we show plots comparing power models. The original Turbostress\textsuperscript{\ref{turbo}} readings are marked on the plots. To these readings, Ichnos+ fits a linear model and a cubic model. We also plot a line that naively assumes linear scaling of power consumption from the readings at 0\% (idle) and 100\% (peak) utilization.
In addition, we plot the memory energy consumption coefficient measured while varying memory load from 0\% (idle) to 100\% (peak) in the second row of the plot.

Figure~\ref{fig:compare-governor-energy-gu} shows that despite both Glasgow cluster nodes having identical hardware resources, their peak CPU power consumption varies substantially. At 100\% utilization and under the same governor settings, Glasgow-15 reached \textasciitilde140$W$ while Glasgow-16 drew only \textasciitilde110$W$, likely attributable to the age of the hardware. This $\approx$27\% difference underscores the importance of node-specific power modeling grounded in actual measurements. 

Figure ~\ref{fig:compare-governor-energy-hu} shows readings for the Berlin cluster nodes. We compare power models generated for four compute nodes, using the governors: performance, powersave, schedutil and ondemand. This newer hardware shows far more consistent behavior, with most readings aligning with other nodes. It also reinforces that the selected governor impacts the power model, with the powersave governor consistently using less power at 100\% utilization. We also observe that the measurements form more non-linear relationships, which highlights why linear models can be less than ideal. 

We calculated the Root Mean Square Error (RMSE) between the model predicted values and the energy consumption readings taken using Turbostress (which uses RAPL to take measurements), for the fitted cubic, fitted linear, and naive linear models. Across all nodes on both clusters, the fitted cubic model demonstrated superior accuracy, markedly outperforming fitted linear which, in turn, proved more accurate than the naive linear model. 

The fitted cubic model proved the most accurate when comparing model predictions and recorded readings. However, when we used the power models to estimate energy consumption for real Nextflow executions and reported the percentage error between the estimated and actual energy consumption of workflows, we found that the fitted linear model \emph{consistently outperformed} the fitted cubic model.
Given these results, we identify the following two drawbacks of using a non-linear model like the fitted cubic model: the impact of background load, and the reliance on coarse-grained CPU utilization data. %

\begin{figure}[bt]
  \centering
  \includegraphics[width=0.8\linewidth,trim={0 0 0 20},clip]{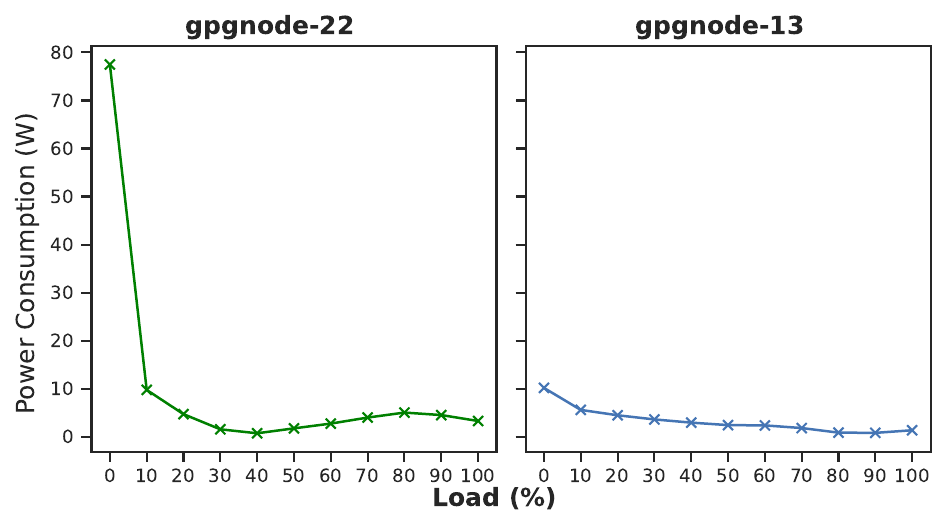}
  \caption{Task power consumption under varying background system loads for Glasgow-22 (left) and Glasgow-13 (right).}
  \label{fig:background-task-impact}
\end{figure}

\subsub{Impact of Background Load}
To enable trace-based resource estimations, we estimate the energy consumption for all individual tasks, considering each task's share of utilized resources in isolation. This works well for several tasks running in parallel on the same shared compute resource only when we use a linear model. 

If we use a cubic model and consider a task that has an average CPU utilization of 100\% on one core, its power consumption will differ depending on whether it runs on its own or shares resources with other CPU-intensive tasks. This effect can be seen in Figure~\ref{fig:background-task-impact}, which shows the power consumption of the commonly used bioinformatics task FastQC\footnote{\url{https://github.com/s-andrews/FastQC}} at different system background loads. This task fully utilizes one core on each system. We can see that the power consumption is significantly affected by the background system load, and this is especially notable for Glasgow-22, which shows a large difference between 0--10\% utilization.

\subsub{Reliance on Coarse-Grained CPU Utilization Data}
Furthermore, the traces generated from Nextflow executions only provide a single coarse-grained CPU utilization average for each individual workflow task -- even for tasks with a runtime spanning seconds to hours and markedly different utilization over time. 

Consequently, we recommend using the generated fitted linear model when estimating energy consumption to reduce the power modelling estimation error -- compared to naive linear and TDP-based methods -- and avoid the estimation being affected by background task utilization and how resource utilization data are aggregated.

\subsection{Energy Consumption Estimates vs. Ground Truth Data}
\label{subsec:vs-rapl}
We executed several Nextflow workflows on three compute clusters and monitored the energy consumed using RAPL. We used Ichnos+ to generate estimations using fitted linear power models. For each cluster, we executed the workflow three times and report the energy measured and estimated for the median workflow run. We report the overall percentage error in the estimation made for the CPU, memory, and overall workflow energy consumption in Table~\ref{table:ichnos-vs-rapl}. 
In the table, we group our results into five groups, each presenting Ichnos+' estimation and the RAPL measurement in kilowatt hours, before presenting the percentage error between them. 

\begin{table*}[]
\caption{Estimated energy consumption from Ichnos+ with fitted power models versus RAPL data (ground truth).} 
\label{table:ichnos-vs-rapl}
\centering
\newcolumntype{g}{>{\columncolor{Blue!20}}c}
\setlength{\tabcolsep}{3pt}
\begin{tabular}{c|ggg|ccc|ggg|ccc|ggg}
\toprule 
& & & & & & & & & & \multicolumn{3}{c}{Total} & \multicolumn{3}{|g}{Total} \\
& \multicolumn{3}{g|}{Node CPU} & \multicolumn{3}{c|}{Task Memory} & \multicolumn{3}{g}{Node Memory} & \multicolumn{3}{|c}{(Node CPU + Task Memory)} & \multicolumn{3}{|g}{(Node CPU + Node Memory)} \\ 
& Ichnos+ & RAPL & Error & Ichnos+ & RAPL & Error & Ichnos+ & RAPL & Error & Ichnos+ & RAPL & Error & Ichnos+ & RAPL & Error \\ 
Workflow & (kWh) & (kWh) & (\%) & (kWh) & (kWh) & (\%) & (kWh) & (kWh) & (\%) & (kWh) & (kWh) & (\%) & (kWh) & (kWh) & (\%)\\
\midrule
\multicolumn{16}{c}{Berlin Cluster} \\
\midrule
Atac-Seq & 0.47 & 0.47 & -0.15 & 0.03 & 0.06 & -52.4 & 0.11 & 0.06 & +79.8 & 0.50 & 0.53 & -5.9 & 0.58 & 0.53 & +9.4 \\
Chip-Seq & 3.53 & 4.12 & -14.4 & 0.30 & 0.54 & -44.1 & 0.53 & 0.54 & -2.5 & 3.83 & 4.66 & -17.9 & 4.06 & 4.66 & -12.9 \\
Nano-Seq & 0.34 & 0.34 & -2.0 & 0.02 & 0.06 & -71.2 & 0.08 & 0.06 & +44.5 & 0.35 & 0.40 & -11.6 & 0.42 & 0.40 & +5.0 \\
Rangeland & 1.56 & 2.00 & -22.3 & 0.17 & 0.25 & -30.5 & 0.26 & 0.25 & +5.6 & 1.73 & 2.25 & -23.2 & 1.82 & 2.25 & -19.1 \\
RNA-Seq & 1.71 & 1.99 & -14.1 & 0.15 & 0.24 & -39.0 & 0.25 & 0.24 & +4.3 & 1.86 & 2.23 & -16.8 & 1.96 & 2.23 & -12.1 \\
Sarek & 3.83 & 4.40 & -12.8 & 0.16 & 0.62 & -74.8 & 0.54 & 0.62 & -12.9 & 3.99 & 5.01 & -20.5 & 4.37 & 5.01 & -12.8 \\
\midrule
\multicolumn{16}{c}{Glasgow Cluster} \\
\midrule
Atac-Seq & 0.92 & 0.9 & +1.96 & 0.12 & 0.17 & -31.06 & 0.18 & 0.17 & +4.0 & 1.04 & 1.08 & -3.37 & 1.10 & 1.08 & +1.85 \\
Chip-Seq & 1.29 & 1.31 & -1.67 & 0.16 & 0.23 & 31.85 & 0.24 & 0.23 & +3.22 & 1.44 & 1.54 & -6.2 & 1.53 & 1.54 & -0.65 \\
Nano-Seq & 0.58 & 0.58 & +0.89 & 0.07 & 0.12 & -44.68 & 0.11 & 0.12 & -10.11 & 0.65 & 0.7 & -7.19 & 0.69 & 0.7 & -1.43 \\
RNA-Seq & 0.99 & 0.96 & +3.06 & 0.12 & 0.18 & -35.99 & 0.19 & 0.18 & +4.44 & 1.11 & 1.15 & -3.16 & 1.18 & 1.15 & +2.61 \\
\midrule
\multicolumn{16}{c}{Dublin Cluster} \\
\midrule
Atac-Seq & 0.66 & 0.71 & -7.06 & 0.1 & 0.27 & -61.06 & 0.23 & 0.27 & -12.98 & 0.76 & 0.97 & -21.85 & 0.89 & 0.97 & -8.25 \\
Chip-Seq & 3.05 & 3.46 & -11.93 & 0.51 & 1.24 & -58.85 & 1.03 & 1.24 & -17.05 & 3.56 & 4.7 & -24.35 & 4.08 & 4.7 & -13.19 \\
Nano-Seq & 0.37 & 0.4 & -6.75 & 0.04 & 0.16 & -74.3 & 0.11 & 0.16 & -25.94 & 0.41 & 0.55 & -25.64 & 0.48 & 0.55 & -12.73 \\
RNA-Seq & 3.53 & 3.98 & -11.34 & 0.66 & 1.56 & -57.85 & 1.34 & 1.56 & -14.02 & 4.19 & 5.54 & -24.42 & 4.87 & 5.54 & -12.09 \\
\bottomrule
\end{tabular}
\end{table*}

Ichnos+ predicts the \emph{node's} overall CPU energy consumption, considering each task's dynamic energy consumption summed with each node's static energy consumption while workflow tasks are running, with a prediction error of $6.9\pm6.5$ (p75 of 12.6, p95 of 17.2).
Ichnos+ predicts the \emph{task} memory energy consumption, considering each task's assigned memory on each node over workflow execution, which does not consider the static energy consumed by memory. This was predicted with an error of $48.6\pm15.5$ (p75 of 60.5, p95 of 74.5). 
Ichnos+ also estimates the \emph{node} memory energy consumption, considering the static energy consumed by the memory of each node used while workflow tasks are running. This had a prediction error of $11.5\pm20.5$ (p75 of 16.3, p95 of 56.8).
Overall, when the predicted \emph{node} CPU energy consumption is summed with the \emph{task} memory energy consumption, and compared to RAPL's total, workflow energy consumption was predicted with an error of $17.3\pm8.3$ (p75 of 22.9, p95 of 24.8). 
Meanwhile, when the predicted \emph{node} CPU energy consumption is summed with \emph{node} memory energy consumption, and compared to RAPL's total, workflow energy consumption was predicted with an error of $10.8\pm5.5$ (p75 of 12.8, p95 of 15.3).

We report both values, distinguishing between task and node memory, as they offer separate advantages. 
Using the task memory allows for task-level estimates to be made, allowing for tasks with significant memory requirements, and therefore energy consumption, to be identified by users.
However, for workflows where some stages of execution have low levels of node memory utilization, like Atac-Seq, Nano-Seq or Sarek, task-level estimates may not capture overall node memory consumption well -- which tends to be constant, leading to significant underestimation.
Here, it may make more sense to use node level memory estimates, which consider the entire node's energy consumption while the overall workflow executes, leading to a reduced prediction error, and overall workflow energy predicted with slightly less variance.

Overall, we observed that the majority of energy consumed by workflow executions was driven by the CPU, which typically consumed four times as much energy as memory.

\subsub{Impact of Background Energy Consumption}
Given that scientific workflows are typically executed on shared infrastructure, or on Kubernetes clusters that require support software to run in the background to facilitate execution, additional energy is consumed. 
Since RAPL measurements are taken at a node-level, we cannot distinguish which processes are responsible for energy consumption without using some attribution-based approach, or attempting to monitor individual processes.
We, hence, investigated the impact of background energy consumption for the Glasgow cluster. We compared our estimations of workflow energy consumption using two power model variations, one generated from measurements with the Kubernetes cluster running and one without. 
Table~\ref{table:ichnos-background-energy-models} shows our estimations made with no load on utilized nodes, and models configured with `background' Kubernetes infrastructure, compared with RAPL readings. All values compared are the mean of three executions.
Given that the power models generated with the Kubernetes cluster running lead to the lowest estimation error, we recommend that users of Ichnos+ should aim to use the most accurate power models available, i.e. training models with background infrastructure, such as Kubernetes clusters already running. If this is not possible, useful estimations can still be made, as estimates are still close to RAPL ground truth data.

\begin{table}
\caption{Ichnos+ energy consumption estimates on the Glasgow Cluster, comparing power models trained with and without background load.}
\label{table:ichnos-background-energy-models}
\centering
\resizebox{0.9\columnwidth}{!}{
\begin{tabular}{c|c|cc|cc}
\toprule 
& RAPL & \multicolumn{2}{c|}{Ichnos+ (no load)} & \multicolumn{2}{c}{Ichnos+ (background)} \\ 
Workload & (kWh) & (kWh) & (\%) & (kWh) & (\%) \\
\midrule 
RNA-Seq & 1.15 & 1.23 & +7.23 & 1.17 & \textbf{+2.16} \\
Atac-Seq & 1.08 & 1.13 & +4.79 & 1.10 & \textbf{+1.83} \\
Chip-Seq & 1.54 & 1.61 & +4.89 & 1.54 & \textbf{-0.04} \\
Nano-Seq & 0.69 & 0.71 & +2.69 & 0.68 & \textbf{-1.82} \\
\bottomrule
\end{tabular}
}
\end{table}

\subsection{Comparison with Baseline Methodology}
\label{subsec:vs-nfco2footprint-plugin}
We compare Ichnos+ predicted workflow energy consumption with the estimation made by the nf-co2footprint plugin\textsuperscript{\ref{foot:nfco2}}, and the measured energy consumption using RAPL. %

In Table~\ref{table:ichnos-vs-baseline}, we show the total energy consumption estimated using Ichnos+ and nf-co2footprint in comparison to RAPL for the median of three executions. We also show the percentage error between these values, with the best performing method highlighted in green and bold. The nf-co2footprint plugin only estimates dynamic energy, without the idle power consumed by CPU and memory. We, hence, manually measured the idle power consumption by reading RAPL counters for each node, and sum it with the dynamic energy for the overall comparison. We also report the average absolute mean error for Ichnos+ and nf-co2footprint for each cluster infrastructure. 

On the Glasgow cluster, Ichnos+ consistently outperforms the plugin which tends to overestimate energy consumption. Ichnos+ mean error is <2\% while nf-co2footprint has a 22\% error. 
On the Berlin cluster, the plugin slightly outperforms Ichnos+ for four workflows, with Ichnos+ significantly outperforming on the other two. We found the Ichnos+ mean error of 12\% remained lower than nf-co2footprint's 17\% error for this cluster. 
Finally, on the Dublin cluster, Ichnos+ consistently outperforms the plugin, which tends to underestimate energy consumption. Ichnos+'s mean error is 12\% while nf-co2footprint has a 35\% error.

We conclude that using an approach that relies solely on the TDP of utilized processors, which do not reflect processor governor settings or variations in the performance of heterogeneous cluster nodes and further does not capture idle power draw, is less accurate than using fitted power models, as Ichnos+ clearly outperforms the nf-core plugin across all three tested cluster environments.

\begin{table}
\caption{The estimated energy consumption using Ichnos+ compared with nf-co2footprint plugin, against RAPL ground truth data. The smaller error is highlighted in bold.}
\label{table:ichnos-vs-baseline}
\centering
\resizebox{0.9\columnwidth}{!}{
\newcolumntype{q}{>{\columncolor{Green!30}}r}
\begin{tabular}{c|c|cc|cc}
\toprule 
& RAPL & \multicolumn{2}{c|}{Ichnos} & \multicolumn{2}{c}{nf-co2footprint} \\ 
Workload & (kWh) & (kWh) & (\%) & (kWh) & (\%) \\
\midrule 
\multicolumn{6}{c}{Glasgow Cluster} \\
\midrule
Atac-Seq & 1.08 & 1.10 & \cellcolor{Green!30}\textbf{{+1.85}} & 1.30 & +20.37 \\
Chip-Seq & 1.54 & 1.54 & \cellcolor{Green!30}\textbf{{0.00}} & 1.92 & +24.68 \\
Nano-Seq & 0.70 & 0.68 & \cellcolor{Green!30}\textbf{{-2.86}} & 0.83 & +18.57 \\
RNA-Seq & 1.15 & 1.18 & \cellcolor{Green!30}\textbf{{+2.61}} & 1.42 & +23.48 \\
\midrule
Average & \cellcolor{Gray!30} & \multicolumn{2}{q|}{\textbf{1.83}} & \multicolumn{2}{r}{21.77} \\
\midrule 
\multicolumn{6}{c}{Berlin Cluster} \\
\midrule
Atac-Seq & 0.53 & 0.58 & \cellcolor{Green!30}\textbf{{+9.43}} & 0.40 & -24.53 \\
Chip-Seq & 4.66 & 4.04 & -13.30 & 4.22 & \cellcolor{Green!30}\textbf{{-9.44}} \\
Nano-Seq & 0.40 & 0.42 & \cellcolor{Green!30}\textbf{{+5.00}} & 0.28 & -30.00 \\
Rangeland & 2.25 & 1.81 & -19.56 & 1.83 & \cellcolor{Green!30}\textbf{{-18.67}} \\
RNA-Seq & 2.23 & 1.96 & -12.11 & 2.05 & \cellcolor{Green!30}\textbf{{-8.07}} \\
Sarek & 5.01 & 4.38 & -12.57 & 4.43 & \cellcolor{Green!30}\textbf{{-11.58}} \\
\midrule
Average & \cellcolor{Gray!30} & \multicolumn{2}{q|}{\textbf{11.99}} & \multicolumn{2}{r}{17.05} \\
\midrule
\multicolumn{6}{c}{Dublin Cluster} \\
\midrule 
Atac-Seq & 0.97 & 0.89 & \cellcolor{Green!30}\textbf{{-8.25}} & 0.66 & -31.96 \\
Chip-Seq & 4.70 & 4.06 & \cellcolor{Green!30}\textbf{{-13.62}} & 3.04 & -35.32 \\
Nano-Seq & 0.55 & 0.48 & \cellcolor{Green!30}\textbf{{-12.73}} & 0.33 & -40.00 \\
RNA-Seq & 5.52 & 4.86 & \cellcolor{Green!30}\textbf{{-11.96}} & 3.70 & -33.06 \\
\midrule
Average & \cellcolor{Gray!30} & \multicolumn{2}{q|}{\textbf{11.64}} & \multicolumn{2}{r}{35.09} \\
\bottomrule 
\end{tabular}
}
\end{table} 

\subsection{Extension of Ichnos+ for Airflow Workflows}
\label{sec:eval:extend}
Ichnos+ requires task-level metrics, but as Airflow does not natively record resource utilization per task, we integrated HyProv~\cite{bountris2025hyprovhybridprovenancemanagement}, a provenance management system capable of capturing execution metrics from the underlying infrastructure. With it, using the relevant APIs, we extracted the hostname, CPU, and memory consumption of each task. We manually captured hardware level information. We then developed a conversion script to map these into the explicit data format required by Ichnos. 

To evaluate the accuracy of the Ichnos+ estimation method when extended to a new system, we developed benchmark workflows on Airflow. The benchmark workflows utilized the sysbench framework\footnote{\url{https://github.com/akopytov/sysbench}}, containerized, structured as four sequential groups of four concurrent tasks. Each task was set to last 10 minutes, yielding a 40-minute total runtime per benchmark workflow. We configured the resource intensity across three distinct tiers (High, Medium, and Low). CPU utilization was enforced using Kubernetes quotas (4.0, 2.0, and 0.5 cores, respectively). To configure memory usage, we managed both the static memory allocation and the active memory traffic. First, we instructed the tasks to allocate a fixed amount of RAM (2048~MB, 1024~MB, and 128~MB) for the duration of the run. Second, to use the memory bus, we configured the tasks to continuously perform random memory writes to this allocated space. Across all configurations, we allowed these write operations to execute continuously as fast as the underlying hardware permitted.

We monitored the real energy consumption using RAPL and compare the estimations with the results in Table~\ref{table:airflow-vs-rapl-data}. 
When using a fitted linear power model, energy consumption is estimated with an error of $13.9 \pm 17.1\%$ for all workloads. We noticed that the workload with low CPU utilization was predicted less well. When this is excluded, we found that the prediction error improves to $13.4 \pm 1.4\%$.

For workloads with low CPU utilization, Ichnos+ significantly overestimates energy consumption when using a fitted linear power model. This aligns with the models plotted for the \texttt{schedutil} governor in Figure~\ref{fig:compare-governor-energy-hu}, where the fitted model overestimates observed power at CPU loads between 0\% and 20\%. Under such low load, employing a na\"ive model, which assumes linear scaling from the minimum to the maximum observed readings, improves the energy consumption estimation ($\approx10\%$ instead of $\approx50\%$), as shown in Table~\ref{table:airflow-vs-rapl-data-minmax}. 

However, we emphasize that these benchmark workloads in Airflow were designed to represent various patterns, but that the extreme case where there is no significant CPU utilization or memory utilization rarely occurs when executing real workflows.
Still, if the utilized workload exhibits such behavior, Ichnos+ with a fitted power model will not be best suited for accurate energy consumption estimation and users might, in these cases, want to opt for a naive linear model. 

\begin{table}
\caption{Estimated energy consumption for benchmark Airflow workloads using Ichnos+ compared to RAPL, I=Ichnos, R=RAPL, E=Error. Selected Fitted--Linear Model for CPU.}
\label{table:airflow-vs-rapl-data}
\centering
\setlength{\tabcolsep}{3pt}
\begin{tabular}{cc|ccc|ccc|ccc}
\toprule 
\multicolumn{2}{c|}{Workload} & \multicolumn{3}{c|}{Node CPU (Wh)} & \multicolumn{3}{c|}{Node Memory (Wh)} & \multicolumn{3}{c}{Total (Wh)} \\
CPU & Mem & I & R & E (\%) & I & R & E (\%) & I & R & E (\%)\\ 
\midrule 
High & High & 210 & 255 & -17.6 & 42 & 45 & -6.7 & 252 & 300 & -16.0 \\
High & Med & 216 & 255 & -15.3 & 43 & 46 & -6.5 & 259 & 301 & -13.9 \\
Med & High & 202 & 235 & -14.0 & 43 & 45 & -4.4 & 245 & 280 & -12.5 \\
Med & Med & 202 & 235 & -14.0 & 43 & 46 & -6.5 & 245 & 281 & -12.8 \\
Low & Low & 190 & 121 & +57.0 & 43 & 28 & +53.6 & 233 & 149 & +56.4 \\
\bottomrule 
\end{tabular}
\end{table} 

\begin{table}
\caption{Estimated energy consumption for benchmark Airflow workloads using Ichnos+ compared to RAPL, I=Ichnos, R=RAPL, E=Error. Selected Naive--Linear Model for CPU.}
\label{table:airflow-vs-rapl-data-minmax}
\centering
\setlength{\tabcolsep}{3pt}
\begin{tabular}{cc|ccc|ccc}
\toprule 
\multicolumn{2}{c|}{Workload} & \multicolumn{3}{c|}{Node CPU (Wh)} & \multicolumn{3}{c}{Total (Wh)} \\
CPU & Mem & I & R & E (\%) & I & R & E (\%)\\ 
\midrule 
\midrule
High & High & 145 & 255 & -43.1 & 187 & 300 & -37.7 \\
Med & Med & 134 & 235 & -43.0 & 177 & 281 & -37.0 \\
Low & Low & 122 & 121 & +0.8 & 165 & 149 & +10.7 \\
\bottomrule 
\end{tabular}
\end{table}

\subsection{Estimation of Environmental Impact} 
\label{sec:eval:impact}
Ichnos+ can be used to estimate the average and marginal carbon emissions as well as embodied carbon emissions, alongside water and land use. The results are shown in Table~\ref{table:env-impact}. The water and land use metrics were configured based on the historical energy mix and the intensity coefficients, as introduced in Section~\ref{subsec:experiment-setup}. We highlight the execution of each workflow on the cluster that minimizes each reported metric.

We retrieved average and marginal CI data aligning with the actual execution times in the regions where compute servers were located. Average CI data was retrieved from Electricity Maps, and Marginal CI data from WattTime\textsuperscript{\ref{foot:watttime}}. Both sources provided data at 5-minute temporal granularity. The estimated carbon emissions produced by Ichnos+ demonstrate that the system enables footprint estimation with the most granular CI data available, while offering users the choice of signal.

We observe that the operational carbon emissions account for the majority of a workflow's carbon footprint, with embodied carbon only responsible for a small fraction. 
We observed that workflows that consumed more energy tended to require more water and land -- values which encompass the water and land effectiveness of the compute nodes, and the water and land used to generate the energy consumed. 

By estimating the carbon emissions of workflow execution, we can explore the impact of carbon-aware shifting and scaling techniques~\cite{west2026systematic}, and their wider influence on the water and land use of workflows. 

\begin{table}
\caption{Estimated average, marginal, and embodied carbon emissions and water and land use. The optimal infrastructure for each impact factor and each workflow is highlighted.}
\label{table:env-impact}
\centering
\begin{tabular}{c|rrr|cc}
\toprule 
& \multicolumn{3}{c|}{Carbon Emissions ($gCO_{2}e$)} & Land & Water \\
Workflow & Average & Marginal & Embodied & Use ($m^2$) & Use (l) \\  
\midrule
\multicolumn{6}{c}{Glasgow Cluster} \\
\midrule
RNA-Seq & \cellcolor{Green!30}290.86 & \cellcolor{green!30}531.86 & \cellcolor{Yellow!30}8.31 & \cellcolor{Brown!30}1.08 & \cellcolor{cyan!30}0.35 \\
Atac-Seq & 251.41 & 496.77 & 7.58 & 1.16 & 0.42 \\
Chip-Seq & \cellcolor{Green!30}393.20 & \cellcolor{green!30}698.19 & \cellcolor{Yellow!30}10.33 & \cellcolor{Brown!30}1.60 & \cellcolor{cyan!30}0.62 \\
Nano-Seq & 146.26 & 314.45 & 4.61 & 0.68 & 0.24 \\
\midrule
\multicolumn{6}{c}{Berlin Cluster} \\ 
\midrule
RNA-Seq & 806.80 & 1,348.30 & 35.93 & 3.78 & 1.75 \\
Chip-Seq & 1,633.32 & 2,976.26 & 62.01 & 7.71 & 3.84 \\
Atac-Seq & 182.65 & 371.67 & 4.16 & \cellcolor{Brown!30}0.40 & \cellcolor{cyan!30}0.16 \\
Rangeland & \cellcolor{Green!30}585.23 & \cellcolor{green!30}1,299.43 & \cellcolor{Yellow!30}52.75 & \cellcolor{Brown!30}2.93 & \cellcolor{cyan!30}1.21 \\
Nano-Seq & 173.24 & 230.19 & 2.55 & \cellcolor{Brown!30}0.24 & \cellcolor{cyan!30}0.14 \\
Sarek & \cellcolor{Green!30}1,861.50 & \cellcolor{green!30}3,096.40 & \cellcolor{Yellow!30}130.18 & \cellcolor{Brown!30}8.38 & \cellcolor{cyan!30}4.06 \\
\midrule
\multicolumn{6}{c}{Dublin Cluster} \\
\midrule
RNA-Seq & 1,061.08 & 2,116.52 & 26.59 & 5.66 & 2.01 \\
Atac-Seq & \cellcolor{Green!30}135.83 & \cellcolor{green!30}370.70 & \cellcolor{Yellow!30}3.50 & 1.17 & 0.39 \\
Chip-Seq & 790.62 & 1,824.27 & 19.39 & 6.07 & 1.95 \\
Nano-Seq & \cellcolor{Green!30}85.13 & \cellcolor{green!30}167.19 & \cellcolor{Yellow!30}1.59 & 0.76 & 0.27 \\
\bottomrule 
\end{tabular}
\end{table}

\section{Limitations} 

\subsub{Generality}
Ichnos+ was created to estimate the carbon footprint of Nextflow workflow executions from trace files. While we additionally exemplify how the system can be extended to support Airflow workflows, we only evaluate the method for two workflow systems, for selected workflows on three infrastructures. There is no confirmation that the system is directly applicable to other workflow systems. However, we detail the requirements to use Ichnos+ in \S\ref{sec:design} and we welcome contributions to our open-source system, which remains under active development. 

\subsub{Power Model Generation and Use}
Ichnos+ requires that resource-specific power models are repeatedly generated when hardware is changed or at regular intervals to track device degradation. However, this is reliant on users having the required permissions available and access to compute infrastructure on which workflows are executed. If power models are not up-to-date or do not align with the original workflow execution, the accuracy of the energy consumption estimations will be limited.
Furthermore, in the rare scenario where workloads have low CPU utilization, using a fitted linear power model may lead to overestimation of CPU energy consumption -- here, it is an option to select a naive linear model for improved accuracy.

\subsub{Limitations of Input Data}
Because Nextflow calculates the average utilization for workflow tasks using their total CPU time, we are able to use Ichnos+ to make estimations of CPU energy consumption with significant accuracy. 
For other workflow systems, it is important that CPU usage is accurately represented, if this is not the case, estimates deviate further from RAPL readings.

Ichnos+ further allows users to provide CI data at varied granularity levels, enabling the use of both average and marginal CI. However, CI data usually specify a value over a given period of time; e.g. WattTime offers marginal CI at intervals of five-minutes, while the National Grid offers average CI at intervals of thirty-minutes. As these intervals become coarser, overall footprint estimation becomes less accurate. Furthermore, we are reliant on these data sources to supply accurate data.

By enabling post-hoc estimation, where the user will likely not have had access to power or emissions monitoring tools, we can only guarantee that our tool uses the estimation methodology described and the user-provided data -- the workflow trace and the CI as well as the power model generated from measurements on compute resources utilized at the time, or as close as possible. This limitation is the same for other existing footprint estimation methodologies like CCF and GA.

\section{Related Work} 
This section examines energy consumption monitoring and modeling methods in a broad context, followed by carbon footprint estimation methodologies and prior research wherein the carbon footprint of scientific workflows has been explored. 

To estimate the carbon footprint of computation, power consumption must first be monitored or modeled. 
Monitoring methods traditionally rely on software interfaces like RAPL -- for Intel Processors -- or the NVIDIA Management Library (NVML) -- for NVIDIA GPUs. Tools built using these interfaces, such as Nvidia-smi, Perf and Scaphandre, can provide accurate measurements of energy consumption~\cite{jay2023experimental}. 
Scaphandre\footnote{https://github.com/hubblo-org/scaphandre} attributes energy to containers based on linear models of CPU and memory utilization. 
Kepler~\cite{amaral_kepler_2023} provides similar container-level attribution, using RAPL-based linear models by default while supporting machine-specific training~\cite{amaral_process-based_2024}.
Both tools integrate with cluster observability stacks such as Prometheus. %
Nf-PEAK~\cite{thamm_nf-peak_2026} enhances RAPL-based attribution on Kubernetes clusters with a non-linear model for CPU and DRAM. It provides a containerized deployment aimed specifically at %
Nextflow workflows. 
However, these methods necessitate configuration prior to workload execution, making them unsuitable for our post-hoc estimation problem space. Instead, we focus on methods capable of retrospectively modeling power consumption based on compute resource usage.

Many methods have been proposed to model server power consumption~\cite{9599719,roy2013energy,dhiman2010system, xiao2013virtual, 10.1145/1250662.1250665, economou2006full, li2012online}.
Some works consider the power consumption of a server to be the sum of idle server consumption -- thought to be a fixed value -- and active consumption caused by computational workloads~\cite{roy2013energy, dhiman2010system, xiao2013virtual}. Other studies use regression models to predict power consumption~\cite{10.1145/1250662.1250665, economou2006full, li2012online}.  
Some works consider the CPU utilization to be the dominant contributor when modelling server consumption (for example, the linear model given by Fan et al.~\cite{10.1145/1250662.1250665}) which produced reasonably accurate estimations and has since been implemented in various estimation methodologies, including CCF. Therefore, we implemented fitted linear-regression models and fitted cubic-regression models for CPU power draw, comparing them with monitored power consumption and existing estimation methodologies.

Several tools have been created to estimate the carbon footprint from computational workloads~\cite{henderson2020systematic, benoit_courty_2024_11171501, eco2AI, anthony2020carbontracker}. Many of these tools model power consumption by using the server utilization and the TDP reported by the manufacturer~\cite{benoit_courty_2024_11171501, eco2AI}. However, this value does not reflect idle power draw and also does not consider processor settings such as the frequency, reducing the estimation's accuracy. Other tools require the user to have privileged (root) access%
~\cite{henderson2020systematic, anthony2020carbontracker, benoit_courty_2024_11171501}. 

Consequently, we focus on existing carbon footprint estimation methodologies that support post-hoc estimation and involve an intermediate step where energy consumption is estimated: CCF\textsuperscript{\ref{ccf-footnote}} and GA~\cite{https://doi.org/10.1002/advs.202100707}. We evaluate Ichnos+ against GA when comparing against the nf-co2footprint plugin\textsuperscript{\ref{foot:nfco2}}.

Prior studies have specifically applied existing carbon footprint estimation methodologies to analyze the footprint of bioinformatics~\cite{10.1093/molbev/msac034}, remote sensing~\cite{thamsen2025energyawareworkflowexecutionoverview} and neuroimaging~\cite{souter2024measuring} research processes. 
In other works where the focus is instead on reducing the energy footprint of scientific workflows, linear power models have been employed to estimate power consumption~\cite{saadiReducingEnergyFootprint2023}. 
The presence of these works clearly indicates interest in being able to estimate the carbon footprint of computation, and validate the use of estimation methodologies in a post-hoc manner.

Beyond carbon emissions, computing exerts significant pressure on water and land resources. 
For example, \cite{li2023} highlight that training large AI models alone can require hundreds of thousands of liters of direct freshwater.
More recently, \cite{jiang2025} proposed a scheduling framework that optimizes carbon and water sustainability across geographically distributed data centers using a MILP-based scheduler.
Similarly, land use has been considered a relevant factor in cloud workload orchestration for workloads such as big data analytics and FaaS~\cite{attenni2025}.
However, the application of such methodologies to scientific computing workflows, to the best of our knowledge, remains largely unexplored. To illustrate the broader environmental footprint of computation beyond carbon, Ichnos+ supports water and land use estimation and we include respective evaluation scenarios.

\subsub{Differences with Prior Work}
\noindent We introduced a first version of Ichnos and presented preliminary results in our previous work~\cite{west2025ichnoscarbonfootprintestimator}. 
In this article, we present Ichnos+, with an expanded methodology with support to estimate embodied carbon emissions, water and land use, using node-specific fitted power models and memory energy coefficients. We significantly expanded our evaluation to measure the accuracy of our energy consumption estimations compared to ground truth RAPL data for distributed workflow executions on three compute infrastructures. We now also compare Ichnos+ with the nf-co2footprint plugin, the accepted tool within the Nextflow nf-core community, which implements the GA methodology. We further demonstrate Ichnos+'s functionality to estimate embodied emissions as well as water and land use.

\section{Conclusion} 
In this paper, we presented Ichnos+, a novel system to estimate the environmental footprint of scientific workflow executions. By analyzing trace files, carbon intensity signals, and hardware-validated power models, Ichnos+ establishes a rigorous system for post-hoc estimation of energy and carbon footprints in data-intensive scientific computing. In addition, the estimated energy footprint can be used to, in turn, estimate additional environmental impacts.

Ichnos+ takes a series of power measurements to generate a fitted linear model used to estimate CPU energy consumption, and measures the memory power consumption to determine the constant used to estimate memory energy consumption. These measurements are repeatable, enabling users to update models when processor settings change or device performance degrades. 
The system extends beyond operational carbon accounting by offering the ability to estimate embodied emissions, water and land use, thereby allowing users to quantify the wider environmental impact of workflow executions. 

Empirical evaluation across diverse compute clusters demonstrates that Ichnos+ predicts energy consumption with a median error of $10.8\pm5.5\%$ compared to hardware-level energy readings using RAPL. 
When compared against the nf-co2footprint plugin (the community-accepted measurement method for Nextflow workflows, which implements the GA methodology), Ichnos+ consistently produced higher accuracy across different infrastructures. 
Finally, we demonstrated the generalizability of our approach by successfully extending Ichnos+ to Apache Airflow. 

\section*{Acknowledgments}
This work was supported by the UK Engineering and Physical Sciences Research Council (EPSRC) under grant number UKRI154, and the German Research Council (DFG) as part of CRC~1404. 
We gratefully acknowledge the sources of electricity grid data: NESO Open Data and Electricity Maps (historical average carbon intensity), and WattTime (marginal operating emission rates).
We thank AWS and Google Cloud for providing research cloud computing credits.

\section*{Data/Code Availability}
An open-source implementation of Ichnos+ is available at {\url{https://github.com/GlasgowC3lab/ichnos}}.

\bibliographystyle{IEEEtran}
\bibliography{references}

\section*{Biographies}
\vspace{-33pt}
\begin{IEEEbiographynophoto}{Kathleen West}
(GS, IEEE) is a PhD student at the University of Glasgow, United Kingdom. She is working on the carbon-aware execution of scientific workflows on heterogeneous clusters. 
\end{IEEEbiographynophoto}
\vspace{-33pt}
\begin{IEEEbiographynophoto}{Youssef Moawad}
is a postdoctoral researcher at the University of Glasgow. He is working on carbon- and performance-aware schedulers for scalable batch processing systems. 
\end{IEEEbiographynophoto}
\vspace{-33pt}
\begin{IEEEbiographynophoto}{Philipp Thamm} is a PhD student in the WBI group at Humboldt-Universität zu Berlin, Germany. His research is focused on energy efficiency, scientific workflows and distributed systems.
\end{IEEEbiographynophoto}
\vspace{-33pt}
\begin{IEEEbiographynophoto}{Vasilis Bountris}
is a PhD student in the WBI group at Humboldt-Universität zu Berlin, Germany. His research is focused on provenance management, scientific workflows and distributed systems.
\end{IEEEbiographynophoto}
\vspace{-33pt}
\begin{IEEEbiographynophoto}{Giulio Attenni}
is a PhD student at Sapienza University of Rome, Italy. His research focuses on applications for sustainable development such as environmentally-conscious cloud orchestration. 
\end{IEEEbiographynophoto}
\vspace{-33pt}
\begin{IEEEbiographynophoto}{Magnus Reid}
earned his MSci degree at the University of Glasgow while his research focused on carbon-aware scientific workflow execution. 
\end{IEEEbiographynophoto}
\vspace{-33pt}
\begin{IEEEbiographynophoto}{Yehia Elkhatib} is an Associate Professor at the University of Glasgow. His research is on data-driven tools to optimize complex distributed systems.
\end{IEEEbiographynophoto}
\vspace{-33pt}
\begin{IEEEbiographynophoto}{Lauritz Thamsen} is a Lecturer / Assistant Professor at the University of Glasgow, where he researches adaptive compute resource management and carbon-aware execution of data-intensive systems.
\end{IEEEbiographynophoto}

\end{document}